\documentclass[aps,twocolumn,superscriptaddress]{revtex4-2}
\usepackage[utf8]{inputenc}
\usepackage{amsmath,bm}
\usepackage{amsfonts}
\usepackage{amssymb}
\usepackage{xcolor}
\usepackage{braket}
\usepackage{graphicx}
\usepackage{subfigure}
\usepackage{subfigure}
\usepackage[colorlinks=true, allcolors=blue]{hyperref}

\begin{document}

\title{Electrically-programmable frequency comb for compact quantum photonic circuits}

\author{Shakir Ullah}
\affiliation{Institute of Nuclear Sciences, Hacettepe University, 06800 Ankara, Turkey}
\affiliation{Department of Physics, Quaid-i-Azam University, Islamabad 45320, Pakistan}

\author{Mehmet Emre Tasgin} \thanks{metasgin@hacettepe.edu.tr}
\affiliation{Institute of Nuclear Sciences, Hacettepe University, 06800 Ankara, Turkey}

\author{Rasim Volga Ovali}
\affiliation{Department of Physics, Recep Tayyip Erdogan University, 53100 Rize, Turkey}

\author{Mehmet G\"{u}nay}
\affiliation{Department of Nanoscience and Nanotechnology, Faculty of Arts and Science, Burdur Mehmet Akif Ersoy University, 15030 Burdur, Turkey}

\date{\today}

\begin{abstract}
Recent efforts have demonstrated the first prototypes for compact and programmable photonic quantum computers~(PQCs). Utilization of time-bin encoding in loop-like architectures enabled programmable generation of quantum states and execution of different~(programmable) logic gates on a single circuit. Actually, there is still space for better compactness and complexity of available quantum states and gate operations: a photonic circuit~(PC) can operate at multiple frequencies. Here, we propose an electrically-programmable frequency comb that generates continuously tunable entanglement among different frequencies. The device is not directly integrated into fragile quantum processing components but is to be used as a fast (picoseconds) tunable auxiliary source provided into state-of-the-art measurement-induced loop-based photonic quantum computers employing programmable (50 MHz) beam-splitters (BSs) and phase-shifters. Multimode entanglement generation is controlled via Fano resonance in the nonlinear response. The generated entanglement can be tuned continuously via an applied voltage which can be delivered to the device via nm-thick wires. The proposed device is integrable, CMOS-compatible, and operates below ps ---but is limited with transistor clock speeds 5-100 GHz. 
\end{abstract}

\maketitle

\section{Introduction}
Demonstrations of quantum supremacy over classical computers~\cite{AWHarrow17,HSZhon20} intensified the research on integrated quantum circuits~(IQCs) where generation, processing, and detection of quantum states can be carried out on a single chip~\cite{GMasada15,VDVaidya20}. This increased the demand of scalable integration of nonlinear optical components, such as squeezers~\cite{FMondain19,YZhang21}, frequency combs~\cite{kues2019quantum} and other nonlinearity generators, into IQC chips. Though integration of various nonlinear components into IQCs are demonstrated successfully~\cite{pelucchi2022potential}, nonlinearity~(for squeezing and entanglement) generation in such devices are either fixed~\cite{YZhao20} or cannot be tuned conveniently~\cite{XLu21,DHallett18,APFoster19,MWang20}.

In order to achieve the quantum supremacy on a wider range of computations, it is necessary to build programmable quantum computers~(QCs)~\cite{madsen2022quantum,asavanant2021time,larsen2021deterministic,takeda2017universal,enomoto2021programmable,arrazola2021quantum,arute2019quantum}, in which flexible quantum states can be generated~\cite{madsen2022quantum,arrazola2021quantum} and various logic gates can be implemented~\cite{takeda2017universal,arrazola2021quantum,larsen2021deterministic,enomoto2021programmable}. In state-of-the-art PQCs, a given initial squeezing~(nonclassicality) is supplied to the circuit on the first module and different entangled states are produced via variable BSs and phase-shifters~\cite{madsen2022quantum,arrazola2021quantum}. In the current PQCs, information is encoded in a pulse train where each pulse corresponds to a mode within the time-domain multiplexing~\cite{yokoyama2013ultra}. The pulses~(modes) are made interact with each other in a loop-structure in order to obtain several-dimensional entangled states. That is, tunability of, for instance, squeezing is rather a new concept~\cite{MGunay23}. Computation can also be performed in similar loop-like architectures where different logic gates can be implemented using measurement-induced squeezing protocol with VBSs~\cite{takeda2017universal,arrazola2021quantum,larsen2021deterministic,enomoto2021programmable}. A single circuit can execute a universal gate~\cite{SLBraunstein05} by winding the input pulse in that programmable circuit. 

Such a loop architecture provides extreme compactness regarding both quantum state preparation and computation. Nevertheless, there is more to do with the compactness and complexity of PQCs~\cite{caspani2016multifrequency}, for example, a given circuit can function at different frequencies~\cite{kues2019quantum}. If one can entangle~(correlate) the pulses of different frequencies~\cite{mahmudlu2023fully}, the functionality of the PQCs can be increased tremendously. For this reason, a close attention is paid on scalable~(integrated) frequency combs~\cite{caspani2016multifrequency,kues2019quantum,kues2017chip,mahmudlu2023fully}. Current electro-optic modulation-based frequency combs can exhibit tunable bandwidth and repetition rate ~\cite{HMurata00,RWu10,KKashiwagi16}. Employing such methods, one can gain partial control also over the nonlinear response of the pulses as the frequency conversion intensity depends on the input intensity of the pulses. Achieving a control merely on the multimode entanglement degree would add an important degree of flexibility on the electrical control quantum optics features --something that has not been demonstrated yet.

Present (not yet programmable) frequency combs can be fabricated using several techniques, e.g., based on a mode-locked laser~\cite{RBWashburn04,SFang13}, electro-optic phase modulation of a laser~\cite{HMurata00,RWu10,KKashiwagi16}, optical micro-resonator~\cite{TJKippenberg11,XYi15,DBraje09}, and cascaded four-wave mixing~(FWM) process~\cite{SJACerqueira08,VSharma21,QLi17}. The frequency comb generators based on FWM process show advantages over the traditional ones~\cite{SJACerqueira08,HZWeng18}. For instance, control of bandwidth and spectral flatness of the frequency combs, generated by phase modulation of a continuous wave, has been demonstrated using cascaded FWM~\cite{VRSupradeepa12}. Moreover, FWM production in a highly nonlinear fiber has also been shown to provide an effective method for obtaining broadband frequency comb generation where the repetition rate tunability can be realized by adjusting the pump wavelength spacing~\cite{EMyslivets12,JFatome06,FCCruz08}. (In these works, enhancement of the cascaded FWM generation is demonstrated by introducing an optical feedback to the input port~\cite{JLi12}.) Hence, here we focus on the control of the FWM process in integrated circuits for programmable frequency combs.

\begin{figure*}
	\centering
	\includegraphics[width=5.8in]{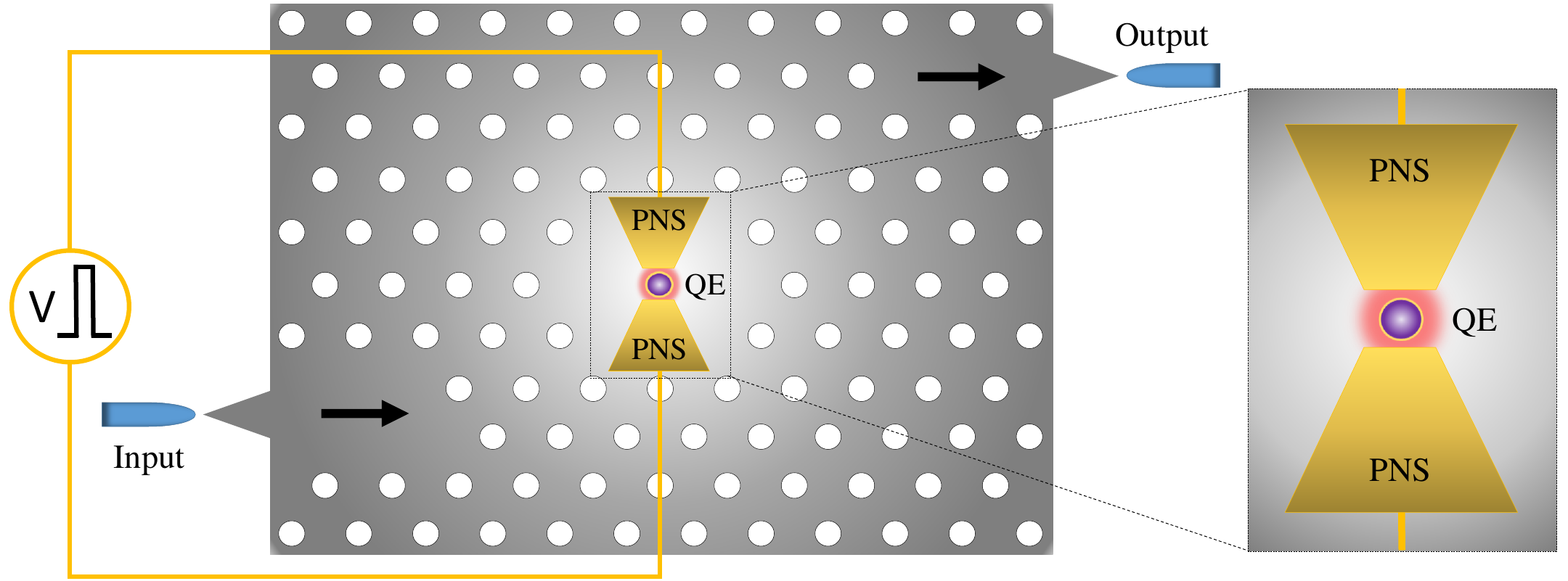}
	\newline
	\caption{Control mechanism of programmable frequency comb which introduces configurable links~(entanglement) among different frequency modes in an IQC. A PNS is placed in a photonic cavity. The resonance of the QE, positioned at the hotspot of the PNS, is controlled electrically via a voltage applied over the PNS itself. The QE resonance controls the FWM process~(see Fig.~\ref{fig4}) taking place within the device.  Cyclic use of the FWM process generates a frequency comb~\cite{kues2019quantum}. Thus, the links among different frequency components can be programmed electrically.}
	\label{fig1}
\end{figure*} 

\begin{figure*}
	\centering
	\includegraphics[width=5.8in]{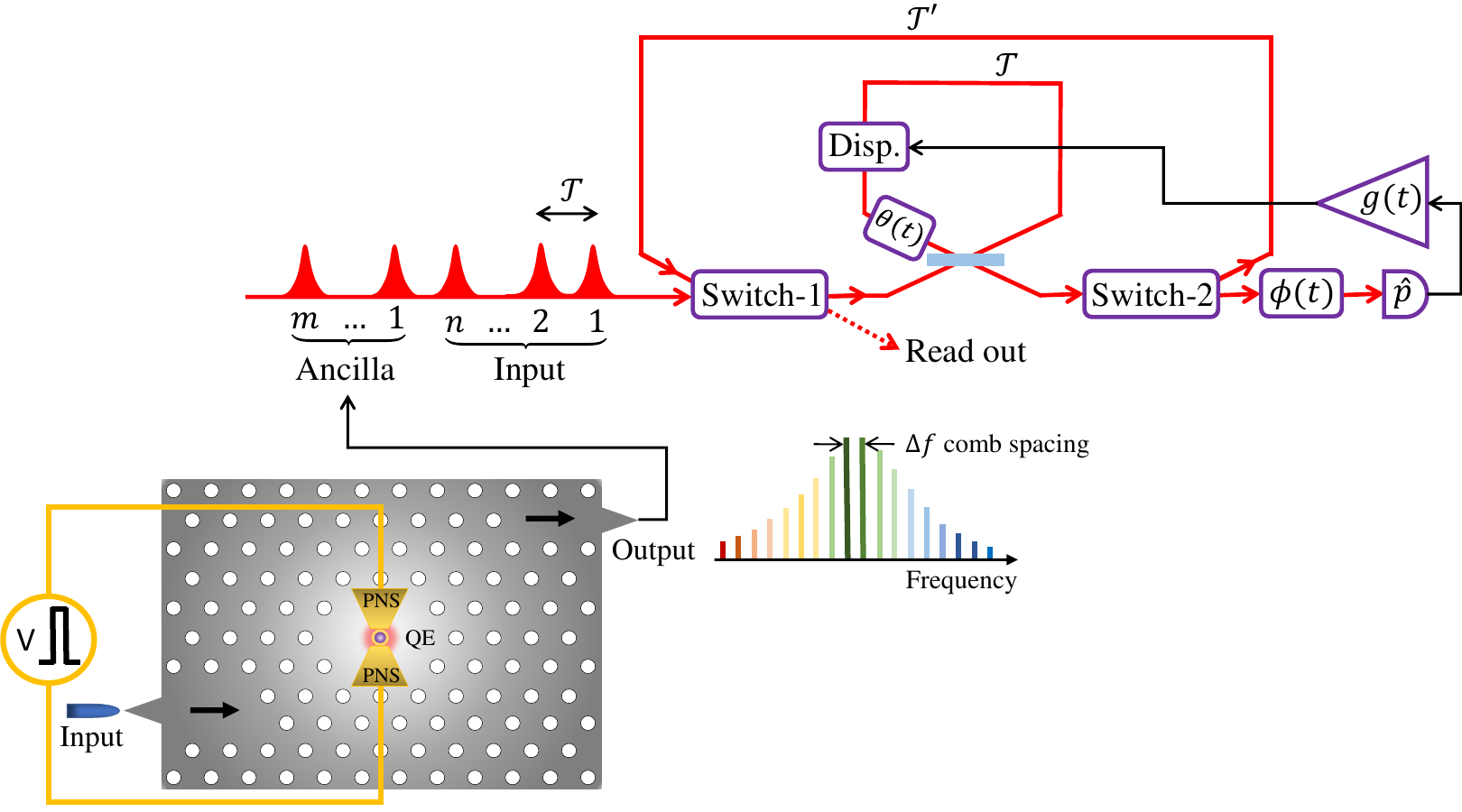}
	\newline
	\caption{Our proposed scheme based on FWM can be used to generate entangled states of different frequencies as the input (as ancilla pulses) to the loop circuit~\cite{takeda2017universal,asavanant2021time,larsen2021deterministic,enomoto2021programmable,enomoto2021programmable}. The loop can support different frequencies operating at the same time.  The frequencies in the output of our auxiliary source is mixed with the quantum states of the same frequencies in the BS.  Measurements (can be at  different times) at the BS outputs integrate the multimode entanglement of the auxiliary source among different frequencies circulating in the same loop. As our device can tune the entanglement degree of the one does not need to employ the programmability of the BSs~(50 MHz) which are slower than our device (5-100 GHz).  }
	\label{fig2}
\end{figure*}

In this paper, we propose a device where the FWM process can be controlled~(continuously tuned) by an applied voltage~\cite{KShibata13,DHallett18,CChakraborty15,SSchwarz16} which can be delivered through nm-thick wires. The device is integrable into quantum photonic circuits and CMOS-compatible. Moreover, it can be used as a fast-tunable device providing auxiliary multi-mode entanglement source into the state-of-the-art loop-based architecture photonic quantum processors~\cite{takeda2017universal,asavanant2021time,larsen2021deterministic}, see the left-bottom panel of Fig.~\ref{fig2}. This protects the fragile quantum states from directly interacting with our metal nanostructure-based device. Moreover, the operation speed of our device, limited with transistor clock speeds 5-100 GHz, is much higher than 50 MHz given in Refs.~\cite{asavanant2021time,larsen2021deterministic}.

A quantum emitter~(QE), located at the hot-spot of a plasmonic nanostructure~(PNS), introduces a Fano resonance in the nonlinear response~\cite{SKSing16,tasgin2018fano}, see Figs.~\ref{fig1} and \ref{fig4}. Depending on the QE's resonance $\Omega_{\rm \scriptscriptstyle QE}$, the FWM process~($\omega_3=1.8\omega_1$) can either be suppressed~(i.e., turned off at $\Omega_{\rm \scriptscriptstyle QE}=\omega_3= 1.8000\omega_1$) or can be further enhanced~(at $\Omega_{\rm \scriptscriptstyle QE}\simeq 1.8005\omega_1$) \textit{on top of} the already-present localization enhancement of the FWM for bare nanostructures, see Fig.~\ref{fig4}. (Here, $\omega_1$ is the pump frequency used in scaling and $\omega_3=1.8\omega_1$ corresponds to the FWM frequency.)

Fano resonances take place at sharp frequency intervals~\cite{postaci2018silent,gunay2020quantum}. This may be disadvantageous for broadband enhancement of nonlinearity. However, here it becomes extremely useful for programming the FWM process via a meV-tuning of the QE's resonance~\cite{KShibata13,DHallett18,CChakraborty15}. One should also note that while plasmonic excitations decay in a few femtoseconds, quantum optics experiments with plasmons~\cite{PhysRevLett.94.110501,tame2013quantum,fasel2006quantum,huck2009demonstration,varro2011hanbury,di2012quantum} clearly show that they can handle quantum features, like squeezing and entanglement, as long as $0.1$ nanoseconds. The latter is determined by noise features~\cite{RSimon94}. 

In the proposed device, the QE resonance is electrically-tuned by an applied voltage, which is delivered to the QE over the nanostructure itself. A few volts is more than sufficient for the tuning  $\Omega_{\rm \scriptscriptstyle QE}= 1.8000$--$1.8005 \times \omega_1$ where $\omega_1$ is typically at the visible or near-IR frequencies. This corresponds to a continuous-control over the FWM intensity, where modulation depths as high as 5 orders of magnitude can be achieved, see Fig.~\ref{fig4}. That is, entanglement among the different frequency modes can be continuously-tuned~(programmed, linked) in the same manner, see Fig.~\ref{fig5}. Similarly, the single-mode nonclassicalities~(e.g., squeezing) of the modes can be continuously-tuned, see Fig.~1 in the Supplementary Material~(SM)~\cite{suppl}. 

Therefore, the proposed device provides an electrically-programmable frequency comb for quantum computers and information processing. The programming ability~(tunability) of the device is continuous. The device provides a  tremendous compactness and extremely enhanced complexity for the information processing due to the existence of  programmable links between different frequency modes.  The links among frequencies can be implemented using the off-line measurement-induced protocols~\cite{filip2005measurement}, once desired multimode entanglement is generated, rather than directly coupling the fragile quantum states to nonlinear media. See left-bottom panel in Fig.~\ref{fig2}. The multi-frequency entanglement source can be implemented into the multi-frequency operating loop similar to the measurement-induced squeezing gate~\cite{takeda2017universal,asavanant2021time,larsen2021deterministic}. Each frequency comprising in the auxiliary source is mixed with fragile quantum state of the loop with the same frequency. When quantum measurements are performed after the BSs, the multi-mode entanglement is integrated among the fragile quantum states belonging to the loop that operate at different frequencies. The degree of the multimode entanglement can be tuned in our auxiliary device without employing electrically programmable BS rotations. The response time of our device is picoseconds and only limited by the transistor clock speeds that can be in order of 5-100 GHz.

Below, we first describe the dynamics of the FWM process taking place in the proposed device where a PNS-QE coupled system is placed into a photonic cavity. (QE can be a quantum dot, defect-center in a 2D material or nanoparticle whose resonance can be electrically tuned.) We obtain Langevin equations governing the dysnamics of plasmon and cavity modes given in Fig.~\ref{fig3}. Before calculating the entanglement among different frequencies, we demonstrate the enhancement of the FWM intensity using c-numbers (semiclassical) at their steady-states, see Fig.~\ref{fig4}. Next, we calculate the entanglement~(Fig.~\ref{fig5}) and  squeezing using the noise operators~\cite{CGenes08,vitali2007optomechanical,gardiner2004quantum,ScullyZubairyBook}. 

\section{Dynamics of the coupled system} 
\begin{figure}
	\centering
	\includegraphics[width=3.3in]{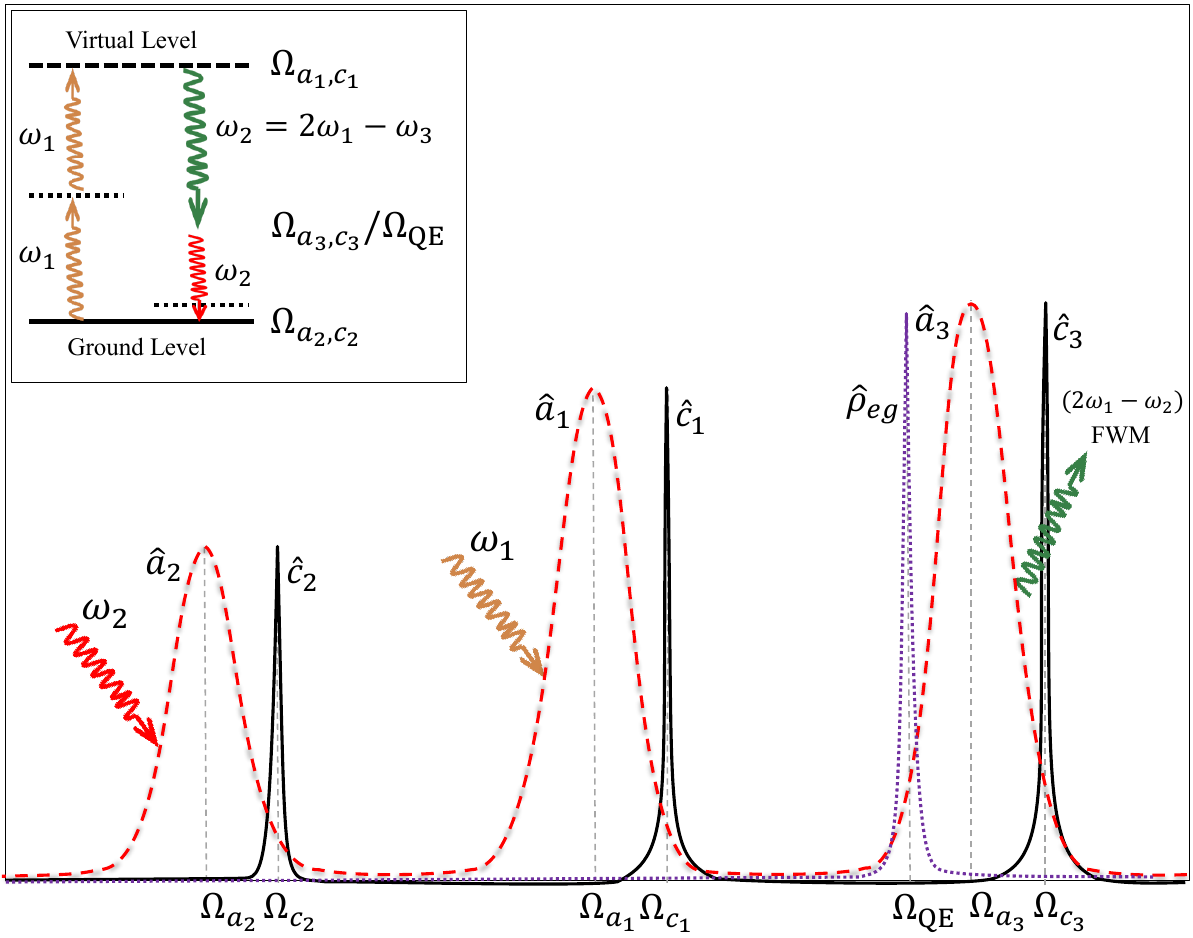}
	\newline
	\caption{Resonances of the cavity ($\hat{c}_{1,2,3}$, $\Omega_{{c}_{1,2,3}}$) and plasmon  ($\hat{a}_{1,2,3}$, $\Omega_{{a}_{1,2,3}}$) modes. Resonance of the QE  $\Omega_{\rm \scriptscriptstyle QE}$ is selected around the $\Omega_{a_3}$ plasmon mode into which FWM oscillations $\omega_3=2\omega_1-\omega_2$ take place. Inset: Schematic of the FWM process $\omega_3=2\omega_1 -\omega_2$. Two pump photons at $\omega_1$ being absorbed to a virtual level, leading to the emission of photons at $\omega_2$ and $\omega_3$ through the third-order nonlinearity $\chi_{ \rm \scriptscriptstyle FWM}$. This diagram corresponds to the FWM interaction terms described in Eqs.~(\ref{eq2})-(\ref{eq6}).}
	\label{fig3}
\end{figure}

The scheme of the device, dynamics of the FWM process and entanglement generation can be described as follows. The PNS, e.g., a bow-tie antenna depicted in Fig.~\ref{fig1}, is placed into a photonic cavity which supports the three related modes~($\hat{c}_{1{\text{--}3}}$, resonances $\Omega_{c_1}$--$\Omega_{c_3}$) for the FWM process, see Fig.~\ref{fig3}. The $\hat{c}_1$  and $\hat{c}_2$ cavity modes are pumped with two integrated lasers of frequencies $\omega_{1}$ and $\omega_{2}$, hamiltonian $\hat{\mathcal{H}}_{\text{las-cav}}$. The cavity fields interact~(strengths $g_{1\text{--}3}$) with the plasmon modes~($\hat{a}_{1\text{--}3}$) of the PNS, see  resonances $\Omega_{a_1}$--$\Omega_{a_3}$ in Fig.~\ref{fig3} respectively; hamiltonian $\hat{\cal{H}}_\text{cav-pls}$. PNS localizes the cavity fields into hot-spots which appears at the centre, e.g., for the bow-tie structure. Localization not only enhances the hot-spot field intensities by orders of magnitude, but also increases the overlap integrals for the nonlinear processes~\cite{SKSing16,tasgin2018fano,ginzburg2012nonlinearly}, here for the FWM. Because of the notorious increase in the overlap integral, nonlinear conversion takes place over plasmons~\cite{grosse2012nonlinear}. The FWM frequency is $\omega_3=2\omega_1-\omega_2$. That is, two plasmons in $\hat{a}_1$ is annihilated and one plasmon  in $\hat{a}_2$ and one plasmon in $\hat{a}_3$ modes are generated, hamiltonian $\hat{\mathcal{H}}_{\rm \scriptscriptstyle FWM}$. 
  
The QE is positioned at the hot-spot of the PNS. QE's resonance, $\Omega_{\rm \scriptscriptstyle QE}$, is chosen such that its coupling~(strength $f$) takes place mainly with the FWM plasmon mode~($\Omega_{a_3}$), hamiltonian $\hat{\cal H}_{\rm pls \text{-}{\scriptscriptstyle QE}}$. (Choice of the values for $\Omega_{\rm \scriptscriptstyle QE}$ in Figs.~\ref{fig4} and \ref{fig5} justifies this notion.)  

Including also energies of the QE, plasmon and cavity modes, hamiltonian $\hat{\mathcal{H}}_{0}$, the total hamiltonian can be written as
\begin{equation}
	\hat{\mathcal{H}}=\hat{\mathcal{H}}_{0} + \hat{\mathcal{H}}_{\text{las-cav}} + \hat{\mathcal{H}}_{\text{cav-pls}} + \hat{\mathcal{H}}_{\rm \scriptscriptstyle FWM} +\hat{\mathcal{H}}_{\text{pls-}{\rm \scriptscriptstyle QE}},
\end{equation}
where
\begin{align}
	&\hat{\mathcal{H}}_{0} = \hbar \Omega_{\rm \scriptscriptstyle QE}\: \hat{\rho}_{ee}+\hbar \sum_{i=1}^{3}(\Omega_{a_i} \hat{a}^{\dagger}_i \hat{a}_i + \Omega_{c_i} \hat{c}^{\dagger}_i \hat{c}_i), \label{eq2}
	\\
	&\hat{\mathcal{H}}_{\text{las-cav}} = i\hbar (\hat{c}^{\dagger}_1 \varepsilon^{(1)}_L \text{e}^{-i \omega_{1} t} + \hat{c}^{\dagger}_2 \varepsilon^{(2)}_L \text{e}^{-i \omega_{2} t} - \text{H.c.}),
	\\
	&\hat{\mathcal{H}}_{\text{cav-pls}} = \hbar (g_1 \hat{c}_1 \hat{a}^\dagger_1 + g_2 \hat{c}_2 \hat{a}^\dagger_2 + g_3 \hat{c}_3 \hat{a}^\dagger_3 + \text{H.c.}),
	\\
	&\hat{\mathcal{H}}_{\rm \scriptscriptstyle FWM} = \hbar \chi_{ \rm \scriptscriptstyle FWM} (\hat{a}^\dagger_3 \hat{a}^\dagger_2 \hat{a}^2_1 + \hat{a}^{\dagger 2}_1 \hat{a}_2 \hat{a}_3),
	\\
	&\hat{\cal H}_{\rm pls \text{-}{\scriptscriptstyle QE}} = \hbar (f \hat{a}^\dagger_3 \hat{\rho}_{ge} + f \hat{a}_3 \hat{\rho}_{eg}). \label{eq6}
\end{align}
Here, $\hat{\rho}_{ee}=|e\rangle\langle e|$,  $\hat{\rho}_{eg}=|e\rangle\langle g|$ and $\hat{\rho}_{ge}=\hat{\rho}_{eg}^\dagger$ represent the density matrices for the QE~\cite{ScullyZubairyBook}.  $\varepsilon^{(1)}_L$ and $\varepsilon^{(2)}_L$ are the pump amplitudes/strengths. $\chi_{\rm \scriptscriptstyle FWM}$ is the overlap integral~\cite{SKSing16,tasgin2018fano,ginzburg2012nonlinearly} for the plasmon modes involving in the FWM process.

{\it Entanglement} is originally generated among the three plasmon modes in the first place due the FWM process inside the cavity. The single-mode nonclassicalities observed in the modes are originally generated in the FWM process. When the FWM Hamiltonian is linearized around the steady-state, it is decomposed into several terms. One term may generate entanglement between the modes and another term (beam-splitter like one) can convert the two-mode entanglement into single-mode squeezing in other modes or simply transfer the single-mode squeezing among the modes. Actually, there exists a conservation relation between single-mode squeezing and entanglement where they can be converted into each other like kinetic and potential energies~\cite{WGe15,LJiru24}. Otherwise, we do not use externally injected squeezed inputs in our calculations. Hence, the plasmon modes interact with the cavity modes via integrated BSs~\cite{LHe20,MMakwana19} with interaction Hamiltonian $(\hat{c}_i^\dagger\hat{a}_i + \text{H.c.})$. Here, the BSs are used only to transfer the inherited squeezing of plasmons modes~\cite{WGe15} into the cavity modes, and due to that the cavity modes are entangled with each other due to entanglement swap mechanism~\cite{sen2005entanglement}. That is, the entanglement initially formed among the plasmon modes are transferred to the cavity modes through their strong coupling. The entanglement is then mapped onto the output optical fields $\hat{c}_{\rm out}^{(1\text{-}3)}$ through the standard input-output interaction, leading to multimode entanglement among the output cavity modes at three distinct frequencies. It is worth to note that, while plasmon-cavity entanglement transfer is affected by the beam-splitter-type coupling, the mapping of intracavity non-classicality onto the output fields depends quantitatively on the input-output interaction and internal losses. Thus, output entanglement (squeezing) may be reduced relative to intracavity values and can be optimized by appropriate coupling.

Recently, programmable and scalable schemes for quantum computing through measurement-induced continuous-variable gates in a loop-based architecture are demonstrated~\cite{takeda2017universal,enomoto2021programmable,enomoto2021programmable}. In the scheme~\cite{takeda2017universal}, a string of \textit{n} pulses of a single spatial mode are used to encode quantum information, which are sent to a nested loop circuit along with the other \textit{m} ancilla pulses. A deterministic squeegeeing quantum gate is then executed in the loop circuit with assistance of the squeezed ancilla pulses by electrically programmable control of optical elements (in particular BS) of the circuit. 
	
The operational capacity of the loop is actually much higher. The loop circuit can operate at different frequencies at the same time. However, this necessitates the programmable entanglement among different frequencies. Our proposed scheme, employing FWM process, can be used to programmably generate multimode entanglement among different frequencies of the comb. This output can be used as an auxiliary multimode entanglement source for the loop-based PQCs~\cite{takeda2017universal,enomoto2021programmable,enomoto2021programmable}, see left-bottom panel of Fig.~\ref{fig2}. In the loop architecture (top panel of Fig.~\ref{fig3}), the switches control routing of the time-bin pulses into and out of the loop, allowing programmable interaction sequences. The dispersion compensators correct pulse broadening and ensure temporal overlap of interacting modes. The two delays $T$ and $T’$ set the timing of successive passes and thereby define the time-bin spacing and nesting required for sequential gate operations. The phase shifter provides programmable phase rotations, while the homodyne detector $\hat{p}$ and the amplifier $g(t)$ are used for measurement-based feedforward (measurement results are scaled by $g(t)$ and applied as displacement operations). Detailed implementation and timing protocols for such loop-based processors are given in Refs.~\cite{takeda2017universal,asavanant2021time,larsen2021deterministic,enomoto2021programmable,enomoto2021programmable}.
	
In loop-architecture PQCs, the squeezing gate is provided by mixing the fragile quantum state with an auxiliary squeezed state at a BS, then performing a measurement at the BS output~\cite{takeda2017universal}. Different amounts of squeezing degrees can be achieved by electrically programming the BS transmission/reflection ratio which takes about 20 ns (50 MHz) to do. We suggest employing a similar approach for integrating our multimode entanglement source into the multi-frequency-operating loop-architecture. Frequencies in the comb are mixed in the BS with the corresponding (same) frequencies in the loop. The measurement at the BS output integrates the multimode entanglement into the circuit, similar to the measurement-induced squeezing gate~\cite{takeda2017universal}. Mixings and measurements of different frequencies can be performed at different times using the same BS and the detector. After the measurements the multimode entanglement is swapped among the different frequency quantum states of the loop. As the degree of the multimode entanglement is tuned by our device, one does not need to tune the BS angles. (Similarly, in our squeezing device~\cite{MGunay23}, squeezing gate can be implemented in much shorter times without tuning the BS angles.)   
	
So, below we study the multimode entanglement features of such a Fano/voltage-controlled fast operating device. We calculate the multimode entanglement for different choices of the applied voltage.

{\it Time-evolution of the modes and QE} can be obtained from the Heisenberg equations of motion, e.g., $i\hbar \dot{\hat{a}}=[\hat{a},\hat{H}]$. Including also the damping/decay rates for the cavity fields~($\kappa_{1\text{--}3}=10^{5}$ Hz), plasmon fields~($\gamma_{1\text{--}3}=10^{14}$ Hz) and the QE~($\gamma_{eg}=10^{9}$ Hz), the Langevin equations can be written as
\begin{align}
	\dot{\hat{c}}_1 =& -(\kappa_1 + i \Omega_{{c}_1})\hat{c}_1 - i g_1^{\ast} \hat{a}_1 + \varepsilon_L^{(1)} \text{e}^{-i \omega_{1}t}, 
	\label{Langevin_c1}
\\
	\dot{\hat{c}}_2 =& -(\kappa_2 + i \Omega_{{c}_2})\hat{c}_2 - i g_2^{\ast} \hat{a}_2 + \varepsilon_L^{(2)} \text{e}^{-i\omega_{2}t}, 
\\
	\dot{\hat{c}}_3 =& -(\kappa_3 + i \Omega_{{c}_3})\hat{c}_3 - ig_3^{\ast} \hat{a}_3, 
\\
	\dot{\hat{a}}_1 =& -(\gamma_1 + i \Omega_{{a}_1})\hat{a}_1 - i g_1 \hat{c}_1 -i 2 \chi_{\rm \scriptscriptstyle FWM} \hat{a}^{\dagger}_1 \hat{a}_2 \hat{a}_3, 
\\
	\dot{\hat{a}}_2 =& -(\gamma_2 + i \Omega_{{a}_2})\hat{a}_2 - i g_2 \hat{c}_2 -i \chi_{\rm \scriptscriptstyle FWM} \hat{a}_3^{\dagger} \hat{a}_1^2, 
\\
	\dot{\hat{a}}_3 = &-(\gamma_3 + i \Omega_{{a}_3})\hat{a}_3 - i g_3 \hat{c}_3 -i f\rho_{ge} -i \chi_{\rm \scriptscriptstyle FWM} \hat{a}_2^{\dagger} \hat{a}_1^2, 
\\
	\dot{\hat{\rho}}_{ge} = &-(\gamma_{eg} + i \Omega_{\rm \scriptscriptstyle QE})\hat{\rho}_{ge} + i f \hat{a}_3 (\hat{\rho}_{gg}-\hat{\rho}_{ee}),
\\
	\dot{\hat{\rho}}_{ee} = &-\gamma_{ee}\hat{\rho}_{ee} + i f \hat{a}_3^{\dagger} \hat{\rho}_{ge}- i f \hat{a}_3 \hat{\rho}_{eg}.
	\label{Langevin_rhoee}
\end{align}
Here, $\gamma_{ee}$ and $\gamma_{eg}$ are the diagonal and off-diagonal elements  decay rates of the QE. We underline that $\chi_{\rm \scriptscriptstyle FWM}$ is considered static and the FWM process is controlled by the applied voltage. The rate Eqs.~(\ref{Langevin_c1})-(\ref{Langevin_rhoee}) reach steady state within a few picoseconds, ensuring stationary operation for all results discussed.

\section{Fano-control of the FWM process} We calculate the entanglement/squeezing by investigating the noise features~(e.g., $\delta \hat{c}$) of the operators around their steady-state values~(c-numbers) $\alpha_c=\langle \hat{c}\rangle$ with $\hat{c}=\alpha+\delta \hat{c}$. Thus, we first calculate the expectations $\alpha$ for the modes and investigate the fluctuations $\delta \hat{c}$ about them. The calculation of the expectations reveals also the Fano-control~(suppression and enhancement) mechanism as follows.

\begin{figure}[t]
	\centering
	\includegraphics[width=3.5in]{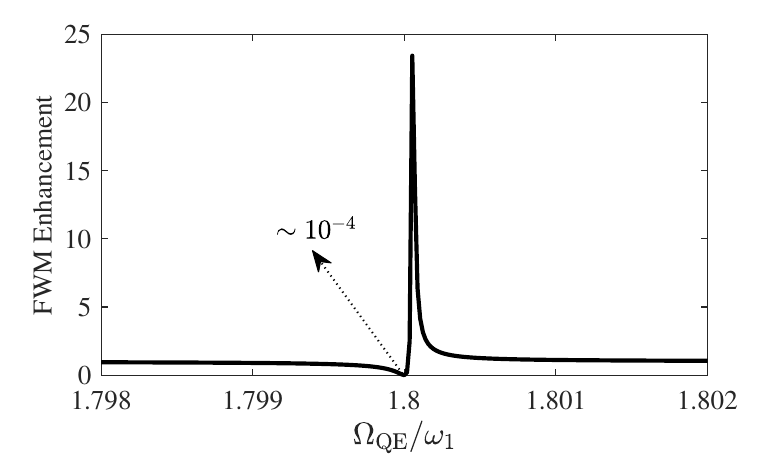}
	\newline
	\caption{Control of FWM process via the choice of the QE resonance. Plotted enhancement factor~(EF) multiplies the EF due to local field enhancement~\cite{renger2009free}. FWM process can be tuned continuously between $10^{-4}$~($\Omega_{\rm \scriptscriptstyle QE}=\omega_3=1.8000\omega_1$) and 25~($\Omega_{\rm \scriptscriptstyle QE}=\omega_3=1.8005\omega_1$) via voltage applied on the QE~\cite{KShibata13,DHallett18,CChakraborty15,SSchwarz16}.}
	\label{fig4}
\end{figure}

We replace the operators in the Langevin equations (\ref{Langevin_c1})-(\ref{Langevin_rhoee}) with their expectations. We determine their steady-state values $\alpha_{c_1 \text{--}c_3}$ and  $\alpha_{a_1 \text{--}a_3}$, see Sec.~2 of SM~\cite{suppl}. For instance, $|\alpha_{a_3}|^2$ gives the number of generated FWM plasmons at the frequency $\omega_3$. As described in Sec. 2 of the SM~\cite{suppl}, an analytical expression for the FWM plasmon amplitude can be obtained as
\begin{equation}
	\alpha_{a_3} = \frac{i \chi_{\rm \scriptscriptstyle FWM} \alpha_{a_2}^{\ast} \alpha_{a_1}^2}{\frac{|f|^2 y}{i(\Omega_{\rm \scriptscriptstyle QE}-\omega_{3}) + \gamma_{eg}} -(i(\Omega_{{a}_3}-\omega_{3})+\gamma_3)},
	\label{Fano_control}
\end{equation}
with $y=\rho_{ee}-\rho_{gg}$ and $\omega_3=2\omega_1-\omega_2$.
While we also calculate the numerical values for $\alpha_{a_3}$ in Fig.~\ref{fig4}, this expression depicts the essential idea below the nonlinearity control. An interference taking place in the denominator of Eq.~(\ref{Fano_control}) is in charge for the enhancement of the FWM.  

The resonance of the QE $\Omega_{\rm \scriptscriptstyle QE}$ can be arranged~(tuned) so that the imaginary part of the first term in the denominator cancels the expression $i(\Omega_{{a}_3}-\omega_3)$ in the second term. This gives the peak in Fig.~\ref{fig4}. We note that enhancement factor~(EF) given in Fig.~\ref{fig4} multiplies the EFs originating due to the field localization enhancements~\cite{SKSing16}.

On the contrary, FWM process can be turned off for the choice of $\Omega_{\rm \scriptscriptstyle QE}=\omega_3$. In this case, the FWM frequency $\omega_3$ production is suppressed. This can be observed from Eq.~(\ref{Fano_control}) as follows.  $\Omega_{\rm \scriptscriptstyle QE}=\omega_3$, the first term of the denominator becomes $|f|^2 y/\gamma_{eg}$ which turns out to be extremely large because of the small decay rate, e.g., $\gamma_{eg}=10^{-6}\omega_{1}$, of the QE~\cite{DHallett18}. We scale frequencies by $\omega_1$. The second term of the denominator is below unity and can be neglected besides $|f|^2 y/\gamma_{eg}$. Thus, the FWM intensity $|\alpha_{a_3}|^2$  is suppressed by orders of magnitude, see the dip~($10^{-4}$) in Fig.~\ref{fig4}. 

\section{Control of entanglement and squeezing}
The links among different frequency modes can also be controlled by the applied voltage. Entanglement and squeezing features are determined by the quantum fluctuations~(quantum noise $\delta \hat{c}_{i}$) about the expectation values $\alpha_{c_i}=\langle \hat{c}_i\rangle$~\cite{gardiner2004quantum}.  We employ a standard method~\cite{CGenes08,vitali2007optomechanical} in our calculations. Equations of motion for the noise operators are obtained by inserting, e.g., the expressions $\hat{c}_i = \alpha_i + \delta \hat{c}_i $, into the Langevin equations~(\ref{Langevin_c1})-(\ref{Langevin_rhoee}), and including the noise operators $\delta \hat{c}_{\text{in}}^{(i)}(t)=-i\sum_{k}\text{e}^{-i\omega_k^{(i)}t} \: \hat{b}_k^{(i)}(0)$, see Sec.~3 of the SM~\cite{suppl}. ($\hat{b}_k$ are vacuum modes, so $\hat{c}_{\text{in}}^{(i)}$ is the vacuum noise~\cite{gardiner2004quantum}.) Ignoring the higher order terms in the equations for the noise operators~\cite{CGenes08,vitali2007optomechanical}, one obtains
\begin{align}
	\delta \dot{\hat{c}}_1 = & -(\kappa_1 + i \Delta_{c_1})\delta \hat{c}_1 - i g_1^{\ast} \delta \hat{a}_1 + \delta \hat{c}_{\text{in}}^{(1)}(t),
	\label{eq:dc1}
\\
	\delta \dot{\hat{c}}_2 =& -(\kappa_2 + i \Delta_{c_2})\delta \hat{c}_2 - i g_2^{\ast} \delta \hat{a}_2 + \delta \hat{c}_{\text{in}}^{(2)}(t),
	\label{eq:dc2}
\\
	\delta \dot{\hat{c}}_3 = & -(\kappa_3 + i \Delta_{c_3})\delta \hat{c}_3 - i g_3^{\ast} \delta \hat{a}_3 ,
	\label{eq:dc3}
\\
	\delta \dot{\hat{a}}_1 = & -(\Gamma_1 + i \Delta_{a_1})\delta \hat{a}_1 - i g_1 \delta \hat{c}_1 -i 2\chi_{\rm \scriptscriptstyle FWM} \nonumber \\
	& \times(\alpha_{a_1}^{\ast}\alpha_{a_2}\delta \hat{a}_3 + \alpha_{a_1}^{\ast}\delta \hat{a}_2 \alpha_{a_3}  +\delta \hat{a}_1^{\dagger}\alpha_{a_2}\alpha_{a_3}),
	\label{eq:da1}
\\
	\delta \dot{\hat{a}}_2 = & -(\Gamma_2 + i \Delta_{a_2})\delta \hat{a}_2 - i g_2 \delta \hat{c}_2 -i\chi_{\rm \scriptscriptstyle FWM} \nonumber \\
	& \times(2\alpha_{a_3}^{\ast}\alpha_{a_1}\delta \hat{a}_1 +\delta \hat{a}_3^{\dagger}\alpha_{a_1}^2 ),
	\label{eq:da2}
\\
	\delta \dot{\hat{a}}_3 = & -(\Gamma_3 + i \Delta_{a_3})\delta \hat{a}_3 - i g_3 \delta \hat{c}_3 -i\chi_{\rm \scriptscriptstyle FWM} \nonumber \\
	& \times(2\alpha_{a_2}^{\ast}\alpha_{a_1}\delta \hat{a}_1 +\delta \hat{a}_2^{\dagger}\alpha_{a_1}^2).
	\label{eq:da3}
\end{align} 
where $\Delta_{c_i}=\omega_{{c}_i}-\omega_{i}$ and $\Delta_{a_i}=\omega_{{a}_i}-\omega_{i}$, with $i=1,2,3$, represent detuning of the cavity and plasmonic modes from the laser fields. Here $\Gamma_{1\text{--}3}$ are the damping rates of the plasmonic noise operators $\delta \hat{a}_{1\text{--}3}$. We note that the same damping rates $\gamma_{1\text{--}3}=\Gamma_{1\text{--}3}=10^{14}$ Hz are used for both the plasmon modes and their fluctuations to maintain consistency with the linearization derived from the Heisenberg-Langevin equations.


We calculate the entanglement among the three output modes of the cavity $\hat{c}_{\text{out}}^{(1\text{--}3)}$ each operating at different frequencies $\omega_{1\text{--}3}$. The output modes are obtained from $\hat{c}_{\text{out}}^{(i)}=r_i \hat{c}_i(t) + \hat{c}_{\text{in}}^{(i)}$ using the input-output formalism~\cite{gardiner2004quantum,ScullyZubairyBook}, where $r_i$ are constants related with the cavity-vacuum coupling rates. See Sec.~3 of the SM~\cite{suppl}.

We use logarithmic negativity~(log-neg, $E_{\cal N}$) for quantifying the entanglement~\cite{KZyczkowski98,GVidal02,GAdesso04,MBPlenio05,STserkis17}. Log-neg is a measure~\cite{MBPlenio05} of entanglement for Gaussian states. The linearization treatment~\cite{CGenes08,vitali2007optomechanical}, we employ here, leaves the quantum states Gaussian,  so that here we can use log-neg as a measure. We also measure the nonclassicality~(e.g., squeezing) of the output modes in units of log-neg. We employ the concept of entanglement-potential~\cite{asboth2005computable}, which weighs the nonclassicality of a mode in terms of the entanglement it generates after a BS~\cite{asboth2005computable,tasgin2020measuring}.

\begin{figure}
	\includegraphics[width=3.5in]{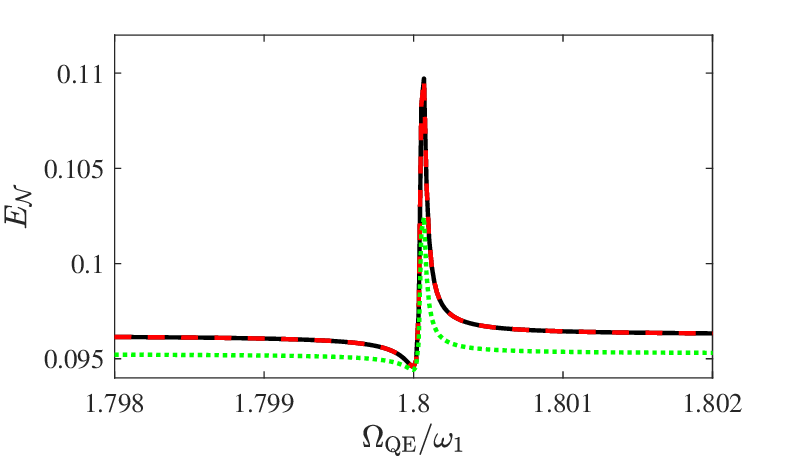}
		\caption{Entanglement in units of log-neg~($E_{\cal N}$)~\cite{KZyczkowski98,GVidal02,GAdesso04,MBPlenio05} between different frequency output modes of the cavity: $\hat{c}_{\text{out}}^{(1)}$--$\hat{c}_{\text{out}}^{(2)}$~(solid black line), $\hat{c}_{\text{out}}^{(1)}$--$\hat{c}_{\text{out}}^{(3)}$~(red dashed line) and $\hat{c}_{\text{out}}^{(2)}$--$\hat{c}_{\text{out}}^{(3)}$~(dotted green line). Entanglement can be continuously tuned by applied voltage which is below a single volt.}
		\label{fig5}
\end{figure}

In Fig.~\ref{fig5}, we present the log-neg entanglement among the three output frequency modes of the cavity, i.e., $\hat{c}_{\text{out}}^{(1)}$-$\hat{c}_{\text{out}}^{(2)}$, $\hat{c}_{\text{out}}^{(2)}$-$\hat{c}_{\text{out}}^{(3)}$, and $\hat{c}_{\text{out}}^{(1)}$-$\hat{c}_{\text{out}}^{(3)}$. Tuning the QE resonance between $\Omega_{\rm \scriptscriptstyle QE}=1.8000\omega_1$ and $1.8005\omega_1$, one can continuously control the entanglement~(link) between different frequencies. The interval  $1.8005-1.8000=0.0005\omega_1$ corresponds to sub-meV tuning for a typical QE. For instance, in the experiment~\cite{KShibata13} 1 Volt can tune a QD about 10 meV and much betters ones are available~\cite{DHallett18,CChakraborty15,SSchwarz16,JMuller05,SAEmpedocles97}. 

The proposed architecture operates in the low-power regime (tens of $\mu W$ optical pump), owing to the strong plasmon-enhanced nonlinearity. Electrical control consumes negligible power, and all relevant noise sources, including cavity and plasmonic losses and quantum fluctuations, are incorporated through the Langevin noise operators, ensuring realistic evaluation of the generated entanglement. We do not put the fragile quantum states generated during the information processing directly into our electrically-programmable multi-mode entanglement device. Multi-mode entangled pulses (reduced noise) generated in our device is coupled to the gate operation through beam-splitter~\cite{takeda2017universal,asavanant2021time,larsen2021deterministic,enomoto2021programmable,enomoto2021programmable}.

From real-world fabrication and integration point of view, the proposed architecture can be implemented using established nanofabrication and integration techniques. Deterministic placement of quantum emitters at plasmonic hot-spots can be achieved via site-controlled growth, transfer-printing or deterministic positioning of emitters~\cite{KShibata13,CChakraborty15,SSchwarz16,JMuller05,SAEmpedocles97}. Nanoscale electrodes for electrical tuning are routinely fabricated using electron-beam lithography~\cite{DHallett18}, and hybrid plasmonic–dielectric designs and post-fabrication tuning can mitigate loss and permit resonance alignment~\cite{YYang17,SVBoriskina17}.
	
The proposed device generates multimode entanglement that can be coupled into the loop-based PQC using conventional beam-splitter or evanescent couplers, similar to the auxiliary squeezed-state injection used in measurement-induced gates~\cite{takeda2017universal}. The device is fully CMOS-compatible and can be integrated with silicon-nitride or lithium-niobate photonic platforms through flip-chip or evanescent coupling techniques~\cite{takeda2017universal,asavanant2021time,larsen2021deterministic,enomoto2021programmable,enomoto2021programmable,IMSoganci13}. The precise positioning of the quantum emitter and formation of nanoscale electrical contacts rely on well-established nanofabrication methods~\cite{DHallett18,KShibata13,CChakraborty15,SSchwarz16,JMuller05,SAEmpedocles97}.

%


%

\section{Summary and outlook}
Recent efforts seek programmable and compact QCs capable of generating and executing configurable quantum states and logic gates. An IQC can function at different frequencies. The recent compactness provided by the employment of time-bin encoding~\cite{yokoyama2013ultra,takeda2017universal,arrazola2021quantum,larsen2021deterministic} can be tremendously enhanced by enabling different frequency modes of the IQC to interact with each other~\cite{caspani2016multifrequency,kues2019quantum,kues2017chip,mahmudlu2023fully}. 
Here, we demonstrate the generation of programmable links among different frequency modes, leading to ultra-compact programmable QCs where links can be configured among time-bin encoded photons (modes) functioning at different frequencies. 
Programmed links among different frequency operations can be established via measurement-induced protocol~\cite{filip2005measurement} without directly coupling the fragile quantum states into nonlinear media. We propose to use our device as an auxiliary  multimode entanglement source that is to be provided into loop-architecture IQCs. This is similar to measurement-induced gate operations where auxiliary squeezed state is implemented into the loop by mixing the source with fragile quantum state operating in the loop~\cite{takeda2017universal}. The device operates at sub-picosecond frequencies (limited solely by the transistor clock speeds 5-100 GHz) and is CMOS-compatible. Thus, the device provides a game-changing increase in the capacity, compactness, and controllability of QCs.

Further work may explore the influence of high-speed electrical modulation on comb stability, improved hybrid plasmonic-dielectric designs to enhance efficiency, and experimental realization of deterministic emitter placement and on-chip integration with continuous-variable processors.

\begin{acknowledgements}
SU, MET, and RVO acknowledge support from T\"{U}B\.{I}TAK-1001 Grant No. 121F141. MET and MG is funded by  T\"{U}B\.{I}TAK-1001 Grant No. 117F118.
\end{acknowledgements}

\section*{Supporting Information Available}
Supplementary material (SM) is provided in which we present further details of our calculations. This material is available free of charge via the Internet at http://pubs.acs.org

\section*{AUTHOR DECLARATIONS}
The authors have no conflicts to disclose.

\section*{DATA AVAILABILITY}
The data that support the findings of this study are available from the corresponding author upon reasonable request.

\bibliography{bibliography}	

@PREAMBLE{
	"\providecommand{\noopsort}[1]{}" 
	# "\providecommand{\singleletter}[1]{#1}%" }

@article{MMakwana19,
	author = {Mehul Makwana and Richard Craster and S\'{e}bastien Guenneau},
	journal = {Opt. Express},
	number = {11},
	pages = {16088--16102},
	publisher = {Optica Publishing Group},
	title = {Topological beam-splitting in photonic crystals},
	volume = {27},
	month = {May},
	year = {2019},
	url = {https://opg.optica.org/oe/abstract.cfm?URI=oe-27-11-16088},
	doi = {10.1364/OE.27.016088},
}

@article{LHe20,
	author = {L. He and H. Y. Ji and Y. J. Wang and X. D. Zhang},
	journal = {Opt. Express},
	number = {23},
	pages = {34015--34023},
	publisher = {Optica Publishing Group},
	title = {Topologically protected beam splitters and logic gates based on two-dimensional silicon photonic crystal slabs},
	volume = {28},
	month = {Nov},
	year = {2020},
	url = {https://opg.optica.org/oe/abstract.cfm?URI=oe-28-23-34015},
	doi = {10.1364/OE.409265},
}

@article{JLi12,
	author = {Jiabao Li and Xiaosheng Xiao and Lingjie Kong and Changxi Yang},
	journal = {Opt. Express},
	number = {20},
	pages = {21940--21945},
	publisher = {Optica Publishing Group},
	title = {Enhancement of cascaded four-wave mixing via optical feedback},
	volume = {20},
	month = {Sep},
	year = {2012},
	url = {https://opg.optica.org/oe/abstract.cfm?URI=oe-20-20-21940},
	doi = {10.1364/OE.20.021940},
}

@article{FCCruz08,
	author = {Flavio C. Cruz},
	journal = {Opt. Express},
	number = {17},
	pages = {13267--13275},
	publisher = {Optica Publishing Group},
	title = {Optical frequency combs generated by fourwave mixing in optical fibers for astrophysical spectrometer calibration and metrology},
	volume = {16},
	month = {Aug},
	year = {2008},
	url = {https://opg.optica.org/oe/abstract.cfm?URI=oe-16-17-13267},
	doi = {10.1364/OE.16.013267},
}

@article{JFatome06,
	author={Fatome, J. and Pitois, S. and Millot, G.},
	journal={IEEE J. Quantum Electron.}, 
	title={20-GHz-to-1-THz Repetition Rate Pulse Sources Based on Multiple Four-Wave Mixing in Optical Fibers}, 
	year={2006},
	volume={42},
	number={10},
	pages={1038-1046},
	doi={10.1109/JQE.2006.881826}
}

@article{HZWeng18,
	author={Weng, Hai-Zhong and Huang, Yong-Zhen and Ma, Xiu-Wen and Yang, Yue-De and Xiao, Jin-Long and Han, Jun-Yuan and Liao, Ming-Long},
	journal={IEEE Photonics J.}, 
	title={Optical Frequency Comb Generation in Highly Nonlinear Fiber With Dual-Mode Square Microlasers}, 
	year={2018},
	volume={10},
	number={2},
	pages={1-9},
	doi={10.1109/JPHOT.2017.2780280}
}

@article{VRSupradeepa12,
	author = {V. R. Supradeepa and Andrew M. Weiner},
	journal = {Opt. Lett.},
	number = {15},
	pages = {3066--3068},
	publisher = {Optica Publishing Group},
	title = {Bandwidth scaling and spectral flatness enhancement of optical frequency combs from phase-modulated continuous-wave lasers using cascaded four-wave mixing},
	volume = {37},
	month = {Aug},
	year = {2012},
	url = {https://opg.optica.org/ol/abstract.cfm?URI=ol-37-15-3066},
	doi = {10.1364/OL.37.003066}
}

@article{QLi17,
	author = {Li, Qing and Jia, Zhi-xu and Li, Zhen-rui and Yang, Yue-de and Xiao, Jin-long and Chen, Shao-wu and Qin, Guan-shi and Huang, Yong-zhen and Qin, Wei-ping},
	title = {Optical frequency combs generated by four-wave mixing in a dual wavelength Brillouin laser cavity},
	journal = {AIP Advances},
	volume = {7},
	number = {7},
	pages = {075215},
	year = {2017},
	month = {07},
	issn = {2158-3226},
	doi = {10.1063/1.4994861},
	url = {https://doi.org/10.1063/1.4994861}
}

@article{VSharma21,
	title = {Demonstration of optical frequency comb generation using four-wave mixing in highly nonlinear fiber},
	journal = {Optik},
	volume = {241},
	pages = {166948},
	year = {2021},
	issn = {0030-4026},
	doi = {https://doi.org/10.1016/j.ijleo.2021.166948},
	url = {https://www.sciencedirect.com/science/article/pii/S0030402621006434},
	author = {Vishal Sharma and Surinder Singh and  Lovkesh and Elena A. Anashkina and Alexey V. Andrianov}
}

@article{EMyslivets12,
	author = {Evgeny Myslivets and Bill P.P. Kuo and Nikola Alic and Stojan Radic},
	journal = {Opt. Express},
	number = {3},
	pages = {3331--3344},
	publisher = {Optica Publishing Group},
	title = {Generation of wideband frequency combs by continuous-wave seeding of multistage mixers with synthesized dispersion},
	volume = {20},
	month = {Jan},
	year = {2012},
	url = {https://opg.optica.org/oe/abstract.cfm?URI=oe-20-3-3331},
	doi = {10.1364/OE.20.003331}
}

@article{SJACerqueira08,
	author = {Arismar Cerqueira S. Jr and J. M. Chavez Boggio and A. A. Rieznik and H. E. Hernandez-Figueroa and H. L. Fragnito and J. C. Knight},
	journal = {Opt. Express},
	number = {4},
	pages = {2816--2828},
	publisher = {Optica Publishing Group},
	title = {Highly efficient generation of broadband cascaded four-wave mixing products},
	volume = {16},
	month = {Feb},
	year = {2008},
	url = {https://opg.optica.org/oe/abstract.cfm?URI=oe-16-4-2816},
	doi = {10.1364/OE.16.002816}
}

@article{XYi15,
	author = {Xu Yi and Qi-Fan Yang and Ki Youl Yang and Myoung-Gyun Suh and Kerry Vahala},
	journal = {Optica},
	number = {12},
	pages = {1078--1085},
	publisher = {Optica Publishing Group},
	title = {Soliton frequency comb at microwave rates in a high-Q silica microresonator},
	volume = {2},
	month = {Dec},
	year = {2015},
	url = {https://opg.optica.org/optica/abstract.cfm?URI=optica-2-12-1078},
	doi = {10.1364/OPTICA.2.001078}
}

@article{DBraje09,
	title = {Brillouin-Enhanced Hyperparametric Generation of an Optical Frequency Comb in a Monolithic Highly Nonlinear Fiber Cavity Pumped by a cw Laser},
	author = {Braje, Danielle and Hollberg, Leo and Diddams, Scott},
	journal = {Phys. Rev. Lett.},
	volume = {102},
	issue = {19},
	pages = {193902},
	numpages = {4},
	year = {2009},
	month = {May},
	publisher = {American Physical Society},
	doi = {10.1103/PhysRevLett.102.193902},
	url = {https://link.aps.org/doi/10.1103/PhysRevLett.102.193902}
}

@article{TJKippenberg11,
	author = {T. J. Kippenberg  and R. Holzwarth  and S. A. Diddams },
	title = {Microresonator-Based Optical Frequency Combs},
	journal = {Science},
	volume = {332},
	number = {6029},
	pages = {555-559},
	year = {2011},
	doi = {10.1126/science.1193968},
	URL = {https://www.science.org/doi/abs/10.1126/science.1193968}
}

@article{KKashiwagi16,
	author = {Ken Kashiwagi and Takashi Kurokawa and Yasushi Okuyama and Takahiro Mori and Yosuke Tanaka and Yoshinori Yamamoto and Masaaki Hirano},
	journal = {Opt. Express},
	number = {8},
	pages = {8120--8131},
	publisher = {Optica Publishing Group},
	title = {Direct generation of 12.5 GHz spaced optical frequency comb with ultrabroad coverage in near-infrared region by cascaded fiber configuration},
	volume = {24},
	month = {Apr},
	year = {2016},
	url = {https://opg.optica.org/oe/fulltext.cfm?uri=oe-24-8-8120&id=338817}
}

@article{RWu10,
	author = {Rui Wu and V. R. Supradeepa and Christopher M. Long and Daniel E. Leaird and Andrew M. Weiner},
	journal = {Opt. Lett.},
	number = {19},
	pages = {3234--3236},
	publisher = {Optica Publishing Group},
	title = {Generation of very flat optical frequency combs from continuous-wave lasers using cascaded intensity and phase modulators driven by tailored radio frequency waveforms},
	volume = {35},
	month = {Oct},
	year = {2010},
	url = {https://opg.optica.org/ol/fulltext.cfm?uri=ol-35-19-3234&id=205870},
}

@article{HMurata00,
	author={Murata, H. and Morimoto, A. and Kobayashi, T. and Yamamoto, S.},
	journal={IEEE J. Sel. Top. Quantum Electron.}, 
	title={Optical pulse generation by electrooptic-modulation method and its application to integrated ultrashort pulse generators}, 
	year={2000},
	volume={6},
	number={6},
	pages={1325--1331},
	url={https://ieeexplore.ieee.org/document/902186}
}

@article{SFang13,
	title={Optical frequency comb with an absolute linewidth of 0.6 Hz--1.2 Hz over an octave spectrum},
	author={Fang, Su and Chen, Haiqin and Wang, Tianyin and Jiang, Yanyi and Bi, Zhiyi and Ma, Longsheng},
	journal={Appl. Phys. Lett.},
	volume={102},
	number={23},
	year={2013},
	publisher={AIP Publishing},
	url={https://pubs.aip.org/aip/apl/article-abstract/102/23/231118/129180/Optical-frequency-comb-with-an-absolute-linewidth?redirectedFrom=fulltext}
}

@article{RBWashburn04,
	author = {Brian R. Washburn and Scott A. Diddams and Nathan R. Newbury and Jeffrey W. Nicholson and Man F. Yan and Carsten G. J{\o}rgensen},
	journal = {Opt. Lett.},
	number = {3},
	pages = {250--252},
	publisher = {Optica Publishing Group},
	title = {Phase-locked, erbium-fiber-laser-based frequency comb in the near infrared},
	volume = {29},
	month = {Feb},
	year = {2004},
	url = {https://opg.optica.org/ol/abstract.cfm?URI=ol-29-3-250},
	doi = {10.1364/OL.29.000250}
}

@article{FMondain19,
	author = {F. Mondain and T. Lunghi and A. Zavatta and E. Gouzien and F. Doutre and others},
	journal = {Photon. Res.},
	number = {7},
	pages = {A36--A39},
	publisher = {Optica Publishing Group},
	title = {Chip-based squeezing at a telecom wavelength},
	volume = {7},
	month = {Jul},
	year = {2019},
	url = {https://opg.optica.org/prj/abstract.cfm?URI=prj-7-7-A36},
	doi = {10.1364/PRJ.7.000A36},
}

@article{YZhang21,
	title={Squeezed light from a nanophotonic molecule},
	author={Zhang, Y and Menotti, M and Tan, K and Vaidya, VD and Mahler, DH and others},
	journal={Nat. Commun.},
	volume={12},
	number={1},
	pages={2233},
	year={2021},
	publisher={Nature Publishing Group UK London},
	url = {https://www.nature.com/articles/s41467-021-22540-2},
}

@article{mahmudlu2023fully,
	title={Fully on-chip photonic turnkey quantum source for entangled qubit/qudit state generation},
	author={Mahmudlu, Hatam and Johanning, Robert and Van Rees, Albert and Khodadad Kashi, Anahita and Epping, J{\"o}rn P and others},
	journal={Nat. Photonics},
	pages={518--524},
	year={2023},
	volume={17},
	url = {https://www.nature.com/articles/s41566-023-01193-1},
	publisher={Nature Publishing Group UK London}
}

@article{pelucchi2022potential,
	title={The potential and global outlook of integrated photonics for quantum technologies},
	author={Pelucchi, Emanuele and Fagas, Giorgos and Aharonovich, Igor and Englund, Dirk and Figueroa, Eden and others},
	journal={Nat. Rev. Phys.},
	volume={4},
	number={3},
	pages={194--208},
	year={2022},
	url = {https://www.nature.com/articles/s42254-021-00398-z},
	publisher={Nature Publishing Group UK London}
}

@article{madsen2022quantum,
	title={Quantum computational advantage with a programmable photonic processor},
	author={Madsen, Lars S and Laudenbach, Fabian and Askarani, Mohsen Falamarzi and Rortais, Fabien and Vincent, Trevor and others},
	journal={Nature},
	volume={606},
	number={7912},
	pages={75--81},
	year={2022},
	url = {https://www.nature.com/articles/s41586-022-04725-x},
	publisher={Nature Publishing Group UK London}
}

@article{enomoto2021programmable,
	title={Programmable and sequential Gaussian gates in a loop-based single-mode photonic quantum processor},
	author={Enomoto, Yutaro and Yonezu, Kazuma and Mitsuhashi, Yosuke and Takase, Kan and Takeda, Shuntaro},
	journal={Sci. Adv.},
	volume={7},
	number={46},
	pages={eabj6624},
	year={2021},
	url = {https://www.science.org/doi/full/10.1126/sciadv.abj6624},
	publisher={American Association for the Advancement of Science}
}

@article{takeda2017universal,
	title={Universal quantum computing with measurement-induced continuous-variable gate sequence in a loop-based architecture},
	author={Takeda, Shuntaro and Furusawa, Akira},
	journal={Phys. Rev. Lett.},
	volume={119},
	number={12},
	pages={120504},
	year={2017},
	url = {https://journals.aps.org/prl/abstract/10.1103/PhysRevLett.119.120504},
	publisher={APS}
}

@article{asavanant2021time,
	title={Time-domain-multiplexed measurement-based quantum operations with 25-MHz clock frequency},
	author={Asavanant, Warit and Charoensombutamon, Baramee and Yokoyama, Shota and Ebihara, Takeru and Nakamura, Tomohiro and others},
	journal={Phys. Rev. Appl.},
	volume={16},
	number={3},
	pages={034005},
	year={2021},
	url = {https://journals.aps.org/prapplied/abstract/10.1103/PhysRevApplied.16.034005},
	publisher={APS}
}

@article{larsen2021deterministic,
	title={Deterministic multi-mode gates on a scalable photonic quantum computing platform},
	author={Larsen, Mikkel V and Guo, Xueshi and Breum, Casper R and Neergaard-Nielsen, Jonas S and Andersen, Ulrik L},
	journal={Nat. Phys.},
	volume={17},
	number={9},
	pages={1018--1023},
	year={2021},
	url = {https://www.nature.com/articles/s41567-021-01296-y},
	publisher={Nature Publishing Group UK London}
}

@article{filip2005measurement,
	title={Measurement-induced continuous-variable quantum interactions},
	author={Filip, Radim and Marek, Petr and Andersen, Ulrik L},
	journal={Phys. Rev. A},
	volume={71},
	number={4},
	pages={042308},
	year={2005},
	url = {https://journals.aps.org/pra/abstract/10.1103/PhysRevA.71.042308},
	publisher={APS}
}

@article{arute2019quantum,
	title={Quantum supremacy using a programmable superconducting processor},
	author={Arute, Frank and Arya, Kunal and Babbush, Ryan and Bacon, Dave and Bardin, Joseph C and Barends, Rami and others},
	journal={Nature},
	volume={574},
	number={7779},
	pages={505--510},
	year={2019},
	url = {https://www.nature.com/articles/s41586-019-1666-5},
	publisher={Nature Publishing Group}
}

@article{caspani2016multifrequency,
	title={Multifrequency sources of quantum correlated photon pairs on-chip: a path toward integrated Quantum Frequency Combs},
	author={Caspani, Lucia and Reimer, Christian and Kues, Michael and Roztocki, Piotr and Clerici, Matteo and others},
	journal={Nanophotonics},
	volume={5},
	number={2},
	pages={351--362},
	year={2016},
	url = {https://www.degruyter.com/document/doi/10.1515/nanoph-2016-0029/html?lang=en},
	publisher={De Gruyter Open}
}

@article{tasgin2018fano,
	title={Fano resonances in the linear and nonlinear plasmonic response},
	author={Tasg{\i}n, Mehmet Emre and Bek, Alpan and Postac{\i}, Selen},
	journal={Fano Resonances in Optics and Microwaves: Physics and Applications},
	volume={219},
	year={2018},
	url = {https://link.springer.com/chapter/10.1007/978-3-319-99731-5_1},
	publisher={Springer}
}

@article{postaci2018silent,
	title={Silent enhancement of SERS signal without increasing hot spot intensities},
	author={Postaci, Selen and Yildiz, Bilge Can and Bek, Alpan and Tasgin, Mehmet Emre},
	journal={Nanophotonics},
	volume={7},
	number={10},
	pages={1687--1695},
	year={2018},
	url = {https://www.degruyter.com/document/doi/10.1515/nanoph-2018-0089/html?lang=en#:~:text=It%20is%20possible%20to%20enhance,the%20Stokes%2Dshifted%20hot%20spots.},
	publisher={De Gruyter}
}

@article{gunay2020quantum,
	title={Quantum emitter interacting with graphene coating in the strong-coupling regime},
	author={G{\"u}nay, Mehmet and Karanikolas, Vasilios and Sahin, Ramazan and Ovali, Rasim Volga and Bek, Alpan and Tasgin, Mehmet Emre},
	journal={Phys. Rev. B},
	volume={101},
	number={16},
	pages={165412},
	year={2020},
	url = {https://journals.aps.org/prb/abstract/10.1103/PhysRevB.101.165412},
	publisher={APS}
}

@article{kues2019quantum,
	title={Quantum optical microcombs},
	author={Kues, Michael and Reimer, Christian and Lukens, Joseph M and Munro, William J and Weiner, Andrew M and Moss, David J and Morandotti, Roberto},
	journal={Nat. Photonics},
	volume={13},
	number={3},
	pages={170--179},
	year={2019},
	url = {https://www.nature.com/articles/s41566-019-0363-0},
	publisher={Nature Publishing Group UK London}
}

@article{kues2017chip,
	title={On-chip generation of high-dimensional entangled quantum states and their coherent control},
	author={Kues, Michael and Reimer, Christian and Roztocki, Piotr and Cort{\'e}s, Luis Romero and Sciara, Stefania and others},
	journal={Nature},
	volume={546},
	number={7660},
	pages={622--626},
	year={2017},
	url = {https://www.nature.com/articles/nature22986},
	publisher={Nature Publishing Group UK London}
}

@article{yokoyama2013ultra,
  title={Ultra-large-scale continuous-variable cluster states multiplexed in the time domain},
  author={Yokoyama, Shota and Ukai, Ryuji and Armstrong, Seiji C and Sornphiphatphong, Chanond and Kaji, Toshiyuki and others},
  journal={Nat. Photonics},
  volume={7},
  number={12},
  pages={982--986},
  year={2013},
  url = {https://www.nature.com/articles/nphoton.2013.287},
  publisher={Nature Publishing Group UK London}
}

@article{arrazola2021quantum,
  title={Quantum circuits with many photons on a programmable nanophotonic chip},
  author={Arrazola, Juan M and Bergholm, Ville and Br{\'a}dler, Kamil and Bromley, Thomas R and Collins, Matt J and others},
  journal={Nature},
  volume={591},
  number={7848},
  pages={54--60},
  year={2021},
  url = {https://www.nature.com/articles/s41586-021-03202-1},
  publisher={Nature Publishing Group UK London}
}

@article{DHallett18,
	author = {Dominic Hallett and Andrew P. Foster and David L. Hurst and Ben Royall and Pieter Kok and others},
	journal = {Optica},
	number = {5},
	pages = {644--650},
	publisher = {Optica Publishing Group},
	title = {Electrical control of nonlinear quantum optics in a nano-photonic waveguide},
	volume = {5},
	month = {May},
	year = {2018},
	url = {http://opg.optica.org/optica/abstract.cfm?URI=optica-5-5-644},
	doi = {10.1364/OPTICA.5.000644}}

@article{KShibata13,
	title={Large modulation of zero-dimensional electronic states in quantum dots by electric-double-layer gating},
	author={Shibata, Kenji and Yuan, Hongtao and Iwasa, Yoshihiro and Hirakawa, Kazuhiko},
	journal={Nature Commun.},
	volume={4},
	number={1},
	pages={1--7},
	year={2013},
	publisher={Nature Publishing Group},
	url = {https://www.nature.com/articles/ncomms3664}}

@article{CChakraborty15,
	title={Voltage-controlled quantum light from an atomically thin semiconductor},
	author={Chakraborty, Chitraleema and Kinnischtzke, Laura and Goodfellow, Kenneth M and Beams, Ryan and Vamivakas, A Nick},
	journal={Nat. Nanotechnol.},
	volume={10},
	number={6},
	pages={507--511},
	year={2015},
	publisher={Nature Publishing Group},
	url = {https://www.nature.com/articles/nnano.2015.79}}

@article{SSchwarz16,
	doi = {10.1088/2053-1583/3/2/025038},
	url = {https://doi.org/10.1088/2053-1583/3/2/025038},
	year = {2016},
	month = {jun},
	publisher = {{IOP} Publishing},
	volume = {3},
	number = {2},
	pages = {025038},
	author = {S Schwarz and A Kozikov and F Withers and J K Maguire and A P Foster and others},
	title = {Electrically pumped single-defect light emitters in WSe2},
	journal = {2D Mater.}}

@article{CGenes08,
	title = {Robust entanglement of a micromechanical resonator with output optical fields},
	author = {Genes, C. and Mari, A. and Tombesi, P. and Vitali, D.},
	journal = {Phys. Rev. A},
	volume = {78},
	issue = {3},
	pages = {032316},
	numpages = {14},
	year = {2008},
	month = {Sep},
	publisher = {American Physical Society},
	doi = {10.1103/PhysRevA.78.032316},
	url = {https://link.aps.org/doi/10.1103/PhysRevA.78.032316}}

@article{PhysRevLett.94.110501,
	title = {Energy-Time Entanglement Preservation in Plasmon-Assisted Light Transmission},
	author = {Fasel, Sylvain and Robin, Franck and Moreno, Esteban and Erni, Daniel and Gisin, Nicolas and Zbinden, Hugo},
	journal = {Phys. Rev. Lett.},
	volume = {94},
	issue = {11},
	pages = {110501},
	numpages = {4},
	year = {2005},
	month = {Mar},
	publisher = {American Physical Society},
	doi = {10.1103/PhysRevLett.94.110501},
	url = {https://link.aps.org/doi/10.1103/PhysRevLett.94.110501}
}

@article{tame2013quantum,
	title={Quantum plasmonics},
	author={Tame, Mark S and McEnery, KR and {\"O}zdemir, {\c{S}}K and Lee, Jinhyoung and Maier, Stefan A and Kim, MS},
	journal={Nat. Phys.},
	volume={9},
	number={6},
	pages={329--340},
	year={2013},
	url = {https://www.nature.com/articles/nphys2615},
	publisher={Nature Publishing Group}
}

@article{fasel2006quantum,
	title={Quantum superposition and entanglement of mesoscopic plasmons},
	author={Fasel, Sylvain and Halder, Matth{\"a}us and Gisin, Nicolas and Zbinden, Hugo},
	journal={New J. Phys.},
	volume={8},
	number={1},
	pages={13},
	year={2006},
	url = {https://iopscience.iop.org/article/10.1088/1367-2630/8/1/013/meta},
	publisher={IOP Publishing}
}

@article{varro2011hanbury,
	title={Hanbury Brown--Twiss type correlations with surface plasmon light},
	author={Varr{\'o}, S{\'a}ndor and Kro{\'o}, Norbert and Oszetzky, D{\'a}niel and Nagy, Attila and Czitrovszky, Alad{\'a}r},
	journal={J. Mod. Opt.},
	volume={58},
	number={21},
	pages={2049--2057},
	year={2011},
	url = {https://arxiv.org/abs/1301.1965},
	publisher={Taylor \& Francis}
}

@article{huck2009demonstration,
	title={Demonstration of quadrature-squeezed surface plasmons in a gold waveguide},
	author={Huck, Alexander and Smolka, Stephan and Lodahl, Peter and S{\o}rensen, Anders S and Boltasseva, Alexandra and Janousek, Jiri and Andersen, Ulrik L},
	journal={Phys. Rev. Lett.},
	volume={102},
	number={24},
	pages={246802},
	year={2009},
	url = {https://journals.aps.org/prl/abstract/10.1103/PhysRevLett.102.246802},
	publisher={APS}
}

@article{di2012quantum,
	title={Quantum statistics of surface plasmon polaritons in metallic stripe waveguides},
	author={Di Martino, Giuliana and Sonnefraud, Yannick and K{\'e}na-Cohen, St{\'e}phane and Tame, Mark and Ozdemir, Sahin K and Kim, MS and Maier, Stefan A},
	journal={Nano Lett.},
	volume={12},
	number={5},
	pages={2504--2508},
	year={2012},
	url = {https://pubs.acs.org/doi/10.1021/nl300671w#},
	publisher={ACS Publications}
}

@book{gardiner2004quantum,
	title={Quantum noise: a handbook of Markovian and non-Markovian quantum stochastic methods with applications to quantum optics},
	author={Gardiner, Crispin and Zoller, Peter and Zoller, Peter},
	year={2004},
	url = {https://link.springer.com/book/9783540223016},
	publisher={Springer Science \& Business Media}
}

@book{ScullyZubairyBook,
	Address = {New York},
	Author = {M. O. Scully and M. S. Zubairy},
	Publisher = {Cambridge University Press},
	Title = {Quantum Optics},
	url = {https://www.cambridge.org/core/books/quantum-optics/08DC53888452CBC6CDC0FD8A1A1A4DD7},
	Year = {1997}}

@article{vitali2007optomechanical,
	title={Optomechanical entanglement between a movable mirror and a cavity field},
	author={Vitali, David and Gigan, Sylvain and Ferreira, Anderson and B{\"o}hm, HR and Tombesi, Paolo and Guerreiro, Ariel and Vedral, Vlatko and Zeilinger, Anton and Aspelmeyer, Markus},
	journal={Phys. Rev. Lett.},
	volume={98},
	number={3},
	pages={030405},
	year={2007},
	url = {https://journals.aps.org/prl/abstract/10.1103/PhysRevLett.98.030405},
	publisher={APS}
}

@article{ginzburg2012nonlinearly,
	title={Nonlinearly coupled localized plasmon resonances: Resonant second-harmonic generation},
	author={Ginzburg, Pavel and Krasavin, Alexey and Sonnefraud, Yannick and Murphy, Antony and Pollard, Robert J and Maier, Stefan A and Zayats, Anatoly V},
	journal={Phys. Rev. B},
	volume={86},
	number={8},
	pages={085422},
	year={2012},
	url = {https://journals.aps.org/prb/abstract/10.1103/PhysRevB.86.085422},
	publisher={APS}
}

@article{grosse2012nonlinear,
	title={Nonlinear plasmon-photon interaction resolved by k-space spectroscopy},
	author={Grosse, Nicolai B and Heckmann, Jan and Woggon, Ulrike},
	journal={Phys. Rev. Lett.},
	volume={108},
	number={13},
	pages={136802},
	year={2012},
	url = {https://journals.aps.org/prl/abstract/10.1103/PhysRevLett.108.136802#:~:text=We%20demonstrate%20how%20k%2Dspace,create%20a%20second%2Dharmonic%20photon.},
	publisher={APS}
}

@article{sen2005entanglement,
	title={Entanglement swapping of noisy states: A kind of superadditivity in nonclassicality},
	author={Sen, Aditi and Sen, Ujjwal and Brukner, {\v{C}}aslav and Bu{\v{z}}ek, Vladim{\'\i}r and {\.Z}ukowski, Marek and others},
	journal={Phys. Rev. A},
	volume={72},
	number={4},
	pages={042310},
	year={2005},
	url = {https://journals.aps.org/pra/abstract/10.1103/PhysRevA.72.042310},
	publisher={APS}
}

@article{renger2009free,
	title={Free-space excitation of propagating surface plasmon polaritons by nonlinear four-wave mixing},
	author={Renger, Jan and Quidant, Romain and Van Hulst, Niek and Palomba, Stefano and Novotny, Lukas},
	journal={Phys. Rev. Lett.},
	volume={103},
	number={26},
	pages={266802},
	year={2009},
	url = {https://journals.aps.org/prl/abstract/10.1103/PhysRevLett.103.266802},
	publisher={APS}
}

@misc{suppl,
	note = {See Supplemental Material}
}

@article{asboth2005computable,
	title={Computable measure of nonclassicality for light},
	author={Asb{\'o}th, J{\'a}nos K and Calsamiglia, John and Ritsch, Helmut},
	journal={Phys. Rev. Lett.},
	volume={94},
	number={17},
	pages={173602},
	year={2005},
	url = {https://journals.aps.org/prl/abstract/10.1103/PhysRevLett.94.173602},
	publisher={APS}
}

@article{tasgin2020measuring,
	title={Measuring nonclassicality of single-mode systems},
	author={Tasgin, Mehmet Emre},
	journal={J. Phys. B: At. Mol. Opt. Phys.},
	volume={53},
	number={17},
	pages={175501},
	year={2020},
	url = {https://iopscience.iop.org/article/10.1088/1361-6455/ab9d02#:~:text=Nonclassicality%20of%20a%20single%2Dmode%20state%20can%20be%20quantified%20with,the%20two%2Dmode%20entanglement%20criteria.},
	publisher={IOP Publishing}
}

@article{AWHarrow17,
	title={Quantum computational supremacy},
	author={Harrow, Aram W and Montanaro, Ashley},
	journal={Nature},
	volume={549},
	number={7671},
	pages={203--209},
	year={2017},
	URL = {https://www.nature.com/articles/nature23458},
	publisher={Nature Publishing Group UK London}
}

@article{HSZhon20,
	author = {Han-Sen Zhong  and Hui Wang  and Yu-Hao Deng  and Ming-Cheng Chen  and Li-Chao Peng  and others},
	title = {Quantum computational advantage using photons},
	journal = {Science},
	volume = {370},
	number = {6523},
	pages = {1460-1463},
	year = {2020},
	doi = {10.1126/science.abe8770},
	URL = {https://www.science.org/doi/abs/10.1126/science.abe8770},
}

@article{VDVaidya20,
	author = {V. D. Vaidya  and B. Morrison  and L. G. Helt  and R. Shahrokshahi  and D. H. Mahler  and others},
	title = {Broadband quadrature-squeezed vacuum and nonclassical photon number correlations from a nanophotonic device},
	journal = {Sci. Adv.},
	volume = {6},
	number = {39},
	pages = {eaba9186},
	year = {2020},
	doi = {10.1126/sciadv.aba9186},
	URL = {https://www.science.org/doi/abs/10.1126/sciadv.aba9186},
}

@article{SLBraunstein05,
	title = {Quantum information with continuous variables},
	author = {Braunstein, Samuel L. and van Loock, Peter},
	journal = {Rev. Mod. Phys.},
	volume = {77},
	issue = {2},
	pages = {513--577},
	numpages = {0},
	year = {2005},
	month = {Jun},
	publisher = {American Physical Society},
	doi = {10.1103/RevModPhys.77.513},
	url = {https://link.aps.org/doi/10.1103/RevModPhys.77.513}
}

@article{GMasada15,
	title={Continuous-variable entanglement on a chip},
	author={Masada, Genta and Miyata, Kazunori and Politi, Alberto and Hashimoto, Toshikazu and O'brien, Jeremy L and Furusawa, Akira},
	journal={Nat. Photonics},
	volume={9},
	number={5},
	pages={316--319},
	year={2015},
	url = {https://www.nature.com/articles/nphoton.2015.42},
	publisher={Nature Publishing Group UK London}
}

@article{YZhao20,
	title = {Near-Degenerate Quadrature-Squeezed Vacuum Generation on a Silicon-Nitride Chip},
	author = {Zhao, Yun and Okawachi, Yoshitomo and Jang, Jae K. and Ji, Xingchen and Lipson, Michal and Gaeta, Alexander L.},
	journal = {Phys. Rev. Lett.},
	volume = {124},
	issue = {19},
	pages = {193601},
	numpages = {7},
	year = {2020},
	month = {May},
	publisher = {American Physical Society},
	doi = {10.1103/PhysRevLett.124.193601},
	url = {https://link.aps.org/doi/10.1103/PhysRevLett.124.193601}
}

@article{XLu21,
	title={Efficient photoinduced second-harmonic generation in silicon nitride photonics},
	author={Lu, Xiyuan and Moille, Gregory and Rao, Ashutosh and Westly, Daron A and Srinivasan, Kartik},
	journal={Nat. Photonics},
	volume={15},
	number={2},
	pages={131--136},
	year={2021},
	publisher={Nature Publishing Group UK London},
	url = {https://www.nature.com/articles/s41566-020-00708-4}
}

@article{APFoster19,
	title = {Tunable Photon Statistics Exploiting the Fano Effect in a Waveguide},
	author = {Foster, A. P. and Hallett, D. and Iorsh, I. V. and Sheldon, S. J. and Godsland, M. R. and others},
	journal = {Phys. Rev. Lett.},
	volume = {122},
	issue = {17},
	pages = {173603},
	numpages = {6},
	year = {2019},
	month = {May},
	publisher = {American Physical Society},
	doi = {10.1103/PhysRevLett.122.173603},
	url = {https://link.aps.org/doi/10.1103/PhysRevLett.122.173603}
}

@article{MWang20,
	doi = {10.1088/1367-2630/ab97ea},
	url = {https://dx.doi.org/10.1088/1367-2630/ab97ea},
	year = {2020},
	month = {jul},
	publisher = {IOP Publishing},
	volume = {22},
	number = {7},
	pages = {073030},
	author = {Min Wang and Ni Yao and Rongbo Wu and Zhiwei Fang and Shilong Lv and Jianhao Zhang and Jintian Lin and Wei Fang and Ya Cheng},
	title = {Strong nonlinear optics in on-chip coupled lithium niobate microdisk photonic molecules},
	journal = {New J. Phys.}
}

@article{MGunay23,
	url = {https://doi.org/10.1515/nanoph-2022-0555},
	title = {On-demand continuous-variable quantum entanglement source for integrated circuits},
	title = {},
	author = {Mehmet Günay and Priyam Das and Emre Yüce and Emre Ozan Polat and Alpan Bek and Mehmet Emre Tasgin},
	pages = {229--237},
	volume = {12},
	number = {2},
	journal = {Nanophotonics},
	doi = {doi:10.1515/nanoph-2022-0555},
	year = {2023},
	lastchecked = {2023-03-24}
}

@article{SKSing16,
	title = {Enhancement of four-wave mixing via interference of multiple plasmonic conversion paths},
	author = {Singh, Shailendra K. and Abak, M. Kurtulus and Tasgin, Mehmet Emre},
	journal = {Phys. Rev. B},
	volume = {93},
	issue = {3},
	pages = {035410},
	numpages = {6},
	year = {2016},
	month = {Jan},
	publisher = {American Physical Society},
	doi = {10.1103/PhysRevB.93.035410},
	url = {https://link.aps.org/doi/10.1103/PhysRevB.93.035410}
}

@article{RSimon94,
	title = {Quantum-noise matrix for multimode systems: U(n) invariance, squeezing, and normal forms},
	author = {Simon, R. and Mukunda, N. and Dutta, Biswadeb},
	journal = {Phys. Rev. A},
	volume = {49},
	issue = {3},
	pages = {1567--1583},
	numpages = {0},
	year = {1994},
	month = {Mar},
	publisher = {American Physical Society},
	doi = {10.1103/PhysRevA.49.1567},
	url = {https://link.aps.org/doi/10.1103/PhysRevA.49.1567}
}

@article{GAdesso04,
	title = {Extremal entanglement and mixedness in continuous variable systems},
	author = {Adesso, Gerardo and Serafini, Alessio and Illuminati, Fabrizio},
	journal = {Phys. Rev. A},
	volume = {70},
	issue = {2},
	pages = {022318},
	numpages = {18},
	year = {2004},
	month = {Aug},
	publisher = {American Physical Society},
	doi = {10.1103/PhysRevA.70.022318},
	url = {https://link.aps.org/doi/10.1103/PhysRevA.70.022318}
}

@article{KZyczkowski98,
	title = {Volume of the set of separable states},
	author = {\ifmmode \dot{Z}\else \.{Z}\fi{}yczkowski, Karol and Horodecki, Pawe\l{} and Sanpera, Anna and Lewenstein, Maciej},
	journal = {Phys. Rev. A},
	volume = {58},
	issue = {2},
	pages = {883--892},
	numpages = {0},
	year = {1998},
	month = {Aug},
	publisher = {American Physical Society},
	doi = {10.1103/PhysRevA.58.883},
	url = {https://link.aps.org/doi/10.1103/PhysRevA.58.883}
}

@article{GVidal02,
	title = {Computable measure of entanglement},
	author = {Vidal, G. and Werner, R. F.},
	journal = {Phys. Rev. A},
	volume = {65},
	issue = {3},
	pages = {032314},
	numpages = {11},
	year = {2002},
	month = {Feb},
	publisher = {American Physical Society},
	doi = {10.1103/PhysRevA.65.032314},
	url = {https://link.aps.org/doi/10.1103/PhysRevA.65.032314}
}

@article{MBPlenio05,
	title = {Logarithmic Negativity: A Full Entanglement Monotone That is not Convex},
	author = {Plenio, M. B.},
	journal = {Phys. Rev. Lett.},
	volume = {95},
	issue = {9},
	pages = {090503},
	numpages = {4},
	year = {2005},
	month = {Aug},
	publisher = {American Physical Society},
	doi = {10.1103/PhysRevLett.95.090503},
	url = {https://link.aps.org/doi/10.1103/PhysRevLett.95.090503}
}

@article{STserkis17,
	title = {Quantifying entanglement in two-mode Gaussian states},
	author = {Tserkis, Spyros and Ralph, Timothy C.},
	journal = {Phys. Rev. A},
	volume = {96},
	issue = {6},
	pages = {062338},
	numpages = {6},
	year = {2017},
	month = {Dec},
	publisher = {American Physical Society},
	doi = {10.1103/PhysRevA.96.062338},
	url = {https://link.aps.org/doi/10.1103/PhysRevA.96.062338}
}

@article{JMuller05,
	title={Wave function engineering in elongated semiconductor nanocrystals with heterogeneous carrier confinement},
	author={M{\"u}ller, J and Lupton, JM and Lagoudakis, PG and Schindler, F and Koeppe, R and Rogach, AL and Feldmann, J and Talapin, DV and Weller, H},
	journal={Nano Lett.},
	volume={5},
	number={10},
	pages={2044--2049},
	year={2005},
	publisher={ACS Publications},
	url = {https://pubs.acs.org/doi/full/10.1021/nl051596x?casa_token=Q53bku81bWkAAAAA%3AwHtdJX9IZCGSq17vodQ6y6Wkc1X0Ks_2TtXX2JXkKQ4d79hNDJMDYHlXm0c6RGlF_xEb9FVRhxy2ziNF#}
}

@article{SAEmpedocles97,
	title={Quantum-confined stark effect in single CdSe nanocrystallite quantum dots},
	author={Empedocles, Stephen A and Bawendi, Moungi G},
	journal={Science},
	volume={278},
	number={5346},
	pages={2114--2117},
	year={1997},
	publisher={American Association for the Advancement of Science},
	url = {https://www.science.org/doi/abs/10.1126/science.278.5346.2114}
}

@article{WGe15,
	title = {Conservation relation of nonclassicality and entanglement for Gaussian states in a beam splitter},
	author = {Ge, Wenchao and Tasgin, Mehmet Emre and Zubairy, M. Suhail},
	journal = {Phys. Rev. A},
	volume = {92},
	issue = {5},
	pages = {052328},
	numpages = {9},
	year = {2015},
	month = {Nov},
	publisher = {American Physical Society},
	doi = {10.1103/PhysRevA.92.052328},
	url = {https://link.aps.org/doi/10.1103/PhysRevA.92.052328}
}

@article{SVBoriskina17,
author = {Svetlana V. Boriskina and Thomas Alan Cooper and Lingping Zeng and George Ni and Jonathan K. Tong and Yoichiro Tsurimaki and Yi Huang and Laureen Meroueh and Gerald Mahan and Gang Chen},
journal = {Adv. Opt. Photon.},
number = {4},
pages = {775--827},
publisher = {Optica Publishing Group},
title = {Losses in plasmonics: from mitigating energy dissipation to embracing loss-enabled functionalities},
volume = {9},
month = {Dec},
year = {2017},
url = {https://opg.optica.org/aop/abstract.cfm?URI=aop-9-4-775},
doi = {10.1364/AOP.9.000775},
}

@article{IMSoganci13,
author = {Ibrahim Murat Soganci and Antonio La Porta and Bert Jan Offrein},
journal = {Opt. Express},
number = {13},
pages = {16075--16085},
publisher = {Optica Publishing Group},
title = {Flip-chip optical couplers with scalable I/O count for silicon photonics},
volume = {21},
month = {Jul},
year = {2013},
url = {https://opg.optica.org/oe/abstract.cfm?URI=oe-21-13-16075},
doi = {10.1364/OE.21.016075},
}

@article{YYang17,
  title={Low-loss plasmonic dielectric nanoresonators},
  author={Yang, Yi and Miller, Owen D and Christensen, Thomas and Joannopoulos, John D and Soljacic, Marin},
  journal={Nano letters},
  volume={17},
  number={5},
  pages={3238--3245},
  year={2017},
url={https://pubs.acs.org/doi/10.1021/acs.nanolett.7b00852},
  publisher={ACS Publications}
}

@article{LJiru24,
  title = {Classical-Nonclassical Polarity of Gaussian States},
  author = {Liu, Jiru and Ge, Wenchao and Zubairy, M. Suhail},
  journal = {Phys. Rev. Lett.},
  volume = {132},
  issue = {24},
  pages = {240201},
  numpages = {7},
  year = {2024},
  month = {Jun},
  publisher = {American Physical Society},
  doi = {10.1103/PhysRevLett.132.240201},
  url = {https://link.aps.org/doi/10.1103/PhysRevLett.132.240201}
}

\end{document}